\documentclass[12p,reviewcopy]{elsart}
\usepackage[sort&compress]{natbib}
\usepackage{graphics}
\usepackage[pdftex,dvips,xdvi]{graphicx}
\usepackage{subfigure}
\usepackage[figuresleft]{rotating}
\usepackage{setspace}
\usepackage{amssymb}
\usepackage{amsmath}
\usepackage{amsfonts}
\usepackage{lscape}

\begin{document}

\begin{frontmatter}

\title{Accounting for outliers and calendar effects in  surrogate simulations of stock return sequences}

\author[Greece]{Alexandros Leontitsis}
\ead{me00743@cc.uoi.gr}
\ead[url]{http://www.geocities.com/CapeCanaveral/Lab/1421/}
\author[Durham]{Constantinos E. Vorlow\corauthref{cor1}}
\ead{Costas@vorlow.org} \ead[url]{http://www.vorlow.org}

\address[Greece]{Department of Education, University of Ioannina,
GR45110, Dourouti, Ioannina, Greece, and Center for Research and
Applications of Nonlinear Systems, University of Patras, GR26500,
Rio, Patras, Greece.} \corauth[cor1]{Corresponding author}

\address[Durham]{Durham Business School, University of Durham, Mill Hill Lane,
Durham, DH13LB, UK. Phone: +44-(0)191-3345727, FAX:
+44-(0)191-3345201}

\date{April 26, 2005}
\journal{ARXIV}

\begin{abstract}

Surrogate Data Analysis (SDA) is a statistical hypothesis testing
framework for the determination of weak chaos in time series
dynamics. Existing SDA procedures do not account properly for the
rich structures observed in stock return sequences, attributed to
the presence of heteroscedasticity, seasonal effects and outliers.
In this paper we suggest a modification of the SDA framework,  based
on the robust estimation of location and scale parameters of
mean-stationary time series and a probabilistic framework which
deals with outliers. A demonstration on the NASDAQ Composite index
daily returns shows that the proposed approach produces surrogates
that faithfully reproduce the structure of the original series while
being manifestations of linear-random dynamics.
\end{abstract}


\end{frontmatter}

\section{Introduction} \label{sec:intro}

The search for nonlinear deterministic dynamics in stock market
prices has been an intensive area\footnote{For early influential
work and discussions refer to
\cite{Bachelier,Granger:64,Samuelson:65,Osborne:62,Fama:65,Fama:70,Mandelbrot63,Mandelbrot67}.}
for research, and especially active in the recent years with the
advances in Econophysics \cite{Ausloos:99,Stanley:00}. The accurate
determination of stock return dynamics and their distributional
properties is of main concern here, as they can significantly
improve portfolio formation and risk evaluation practices, as well
as allow the fine tuning of asset valuation procedures.

There have been several indications that stock prices do not
fluctuate as randomly as they should, according to the underlying
theoretical equilibrium framework \cite[see discussions in Ref.
][]{Ausloos:98,Mand99,Multifract:99,Missbe}, and exhibit rich and
complex structures \cite{KYRTSOU:02,KYRTSOU:03,Vorlow:04a}. However,
earlier research
\cite{Hsieh:91,MM92,BHL92,Pilarinu93,Malliaris:99a,Malliaris:99b}
has not provided a clear  answer towards the presence or absence of
nonlinear determinism and chaos. Hence, the candidacy of
deterministic chaos as an alternative hypothesis to randomness, has
not enjoyed popularity among the ranks of economists. Limitations
posed by the quantity and quality of data, computational power and
the absence of a widely acceptable and appropriate theoretical and
statistical framework, have also been factors that contributed to
the dispute against chaotic dynamics in finance and economics
\cite{Theiler86,Theiler90a,Ramsey:89,Ramsey:90a,Ramsey:90b,Hsieh:91,MM92,BHL92,Granger94chaos}.
However, a Monte-Carlo simulation-based statistical hypothesis
testing framework for  detecting  weak chaos,  appears to have been
ignored by and large till recently in finance
\cite{Kugiumtzis02,Small:03,Soofi:03,Leontitsis:03,Vorlow:04}. This
framework is called Surrogate Data Analysis
\citep[SDA:][]{Theiler88a,Theiler92} and has preceded historically a
significant amount of influential research of chaotic dynamics in
finance and economics.

SDA (see section \ref{sec:sec3} for details on how this methodology
works) has been primarily designed to ensure the validity of results
of investigations for nonlinear determinism and the presence of weak
chaos. Similar investigations have been mainly focusing on the
examination of invariant measures, such as dimension based
statistics for the characterization of strange attractors
\cite{Grassberger83d}. However such measures have been shown to
provide misleading, biased or inconclusive results, due to the
presence of noise or the lack of sufficient observations in the data
sets examined
\cite{Theiler86,Theiler90a,Ramsey:89,Ramsey:90a,Ramsey:90b}. Though
SDA can provide the means to bypass some of the limitations posed by
the quality and quantity of the sequences under examination, still
the structure of their underlying dynamics and their noise content
can pose serious considerations. The above discussion comes into
context in the analysis of financial time series, where the nature
of the data generating processes and the noise components are still
largely unknown, while the ``mechanics" and the equilibrium
conditions of the market systems examined often appear empirically
to be ill or loosely defined. Especially, the presence of
heteroscedastic noise in stock returns and their nonstationary
fluctuations among other stylized facts \cite{Rama01}, can mask the
presence of low dimensional nonlinear determinism. As mentioned
earlier, the greatest disadvantage of the nonlinear statistics based
on invariant measures is their lack of power, especially in
financial applications. SDA enables us to bypass this limitation.
However, heteroscedasticity may render this exercise useless, as the
existing surrogate methods are designed for homoscedastic time
series. Their application on noisy and heteroscedastic sequences may
lead to misleading results and biased or inaccurate conclusions
\cite{Rapp01,Kugium01,Rapp:94}. Since SDA is essentially a
simulation of the linear characteristics of a time series, it should
be able to deal with heteroscedasticity, outliers and calendar
effects, which are major features of financial time series. In this
paper we demonstrate how to modify one of the most advanced and
popular surrogate methods, the Iterative Amplitude Adjusted Fourier
Transformed surrogates (IAAFT) \cite{IAAFT}, in order to account for
important stylized facts regarding heteroscedasticity, calendar
effects and outliers in stock returns sequences.

\section{Dealing with heteroscedasticity and outliers} \label{sec:sec2}

A  time series sequence is subject to heteroscedasticity when the
variance is  time-varying. Empirical research on stock returns has
shown that from time to time the variance fluctuates, volatility
appears to be clustering, while outliers appear in the time series,
often attributed to exogenous factors and random events. The use of
robust statistics is justified  for the identification and
characterization of the underlying dynamics. Robust statistics were
developed principally during the 70's with a few related but major
methodologies appearing the following decade. In this paper we make
use of the Least Median of Squares (LMS) concept introduced by
\cite{Rousseeuw:84,Rousseeuw:87}. The LMS estimator minimizes the
median of the squared discrepancies rather than the mean, as in the
Ordinary Least Squares (OLS) methodology. Hence,  LMS estimators may
produce results which are relatively immune to the presence of the
outliers and the non-normality of the errors' distribution. One
disadvantage of LMS estimators is that they are considerably less
efficient in the case of normally distributed errors. However, it is
well established that the distributions of the first logarithmic
differences of stock prices (i.e., logarithmic returns) fail
normality tests, and exhibit strong leptokurtic features, and this
justifies the applicability of the LMS concept.

Since outliers may pose considerations under the SDA framework, it
is necessary to follow a policy for their classification and
characterization. For the purposes of our approach here, in order to
isolate the outliers of a given data set $x$, we suggest the
following steps in the spirit of the wider LMS literature:

\begin{enumerate}
   \item Find the LMS location parameter of the data set:
   \begin{equation}\label{eq:eq1}
       \mathrm{loc}=\mathrm{argmin}(\mathrm{med}(x-\theta)^2),
   \end{equation} i.e., determine the value of a parameter $\theta$
   which minimizes the median of the squared deviations from the
    median. This can be easily achieved by sorting the data set and
    calculating the midpoint of the range of the 50\% of the densest data.
   \item Find the LMS scale parameter of the data set:
   \begin{equation}\label{eq:eq2}
       \mathrm{scale} = 1.4826 \times ( 1 + \frac{5}{N-1} ) \times \mathrm{med}
       (r^2),
   \end{equation}
    where $r$ is the residuals' vector obtained from the
    previous step and the consistency constant 1.4826 comes from the square root of the median of the chi-square distribution with one degree of freedom \cite{Rousseeuw:84}. Hence, this scale parameter can  be calculated once the
   LMS location parameter $\theta$ is estimated.
   \item Calculate the $\mathrm{z}_{ \mathrm{LMS} }$-score: $\mathrm{z}_{\mathrm{LMS}}=(x-\mathrm{loc})/\mathrm{scale}$ i.e., normalize
   the data according to the LMS concept.
\end{enumerate}

\begin{landscape}
\begin{figure}[ht]
  \centering
  \includegraphics[width=6in]{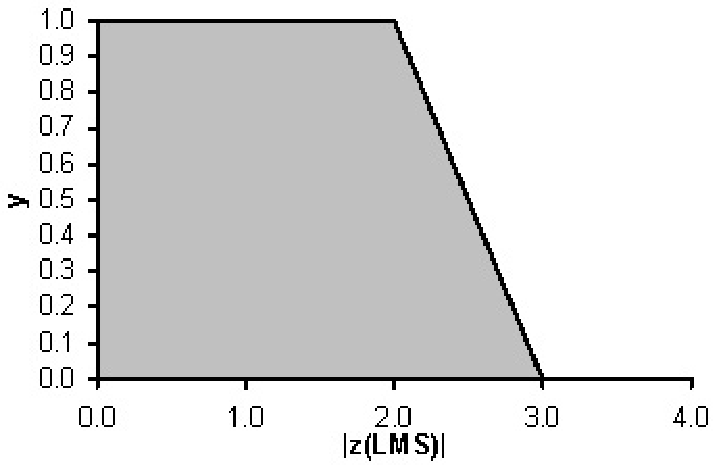}\\
  \caption{The model proposed by Rousseeuw and Leroy (1987) \cite{Rousseeuw:87}
  regarding the distinction between outliers and the  bulk of the observations,
  according to the $|\mathrm{z}_{ \mathrm{LMS}}| $ score. In this model $\lambda$ on the vertical scale  represents the degree of a point
  not being an outlier. Observations with $|\mathrm{z}_{ \mathrm{LMS}}| <1$ are not considered outliers,
  and observations with $|\mathrm{z}_{ \mathrm{LMS} }|>3$ are surely considered outliers.
  In between these two extremes, the degree falls linearly.}\label{fig:fig1}
\end{figure}
\end{landscape}

Rousseeuw and Leroy \cite{Rousseeuw:87} propose the following fuzzy
model (see also Fig. \ref{fig:fig1}) for determining the degree
$\lambda$ of a residual not being an outlier:
\begin{itemize}
\item If $| \mathrm{z}_{\mathrm{LMS}} | \leq 2.0 $ then $\lambda=1.0$ and $x$ is not an
outlier,
\item  if $ 2.0 < | \mathrm{z}_{\mathrm{LMS}} | \leq 3.0$ then  $\lambda=3.0-| \mathrm{z}_{\mathrm{LMS}} | $, and $x$ is not an outlier with degree $\lambda$, and
\item  if $3.0 < | \mathrm{z}_{\mathrm{LMS}} |$ then  $\lambda=0.0$, and
$x$ is an outlier.
\end{itemize}

\medskip
{\begin{center}  [ Insert Fig. \ref{fig:fig1} about here. ]
\end{center}}
\medskip

Our approach converts the above fuzzy model to probabilistic. In
other words, every time we run the surrogate data algorithm we
consider a probability equal to the degree $\lambda$ that a data
point $x$ is classified or not as an outlier. Thus, we classify as
``outliers" values with a corresponding $|\mathrm{z}_{ \mathrm{LMS}
}|$ score more than 3.0, and as non-outliers the values with a
corresponding $|\mathrm{z}_{ \mathrm{LMS} }|$ score less than 2.0. A
random number generator that produces uniformly distributed random
values in [0,1]  helps on the intermediate $|\mathrm{z}_{
\mathrm{LMS} }|$ scores (i.e., scores between 2 and 3). For example
if a data point $x$ has $|\mathrm{z}_{ \mathrm{LMS} }|$ score of
2.8, a corresponding random number of 0.2 or greater will classify
it as an outlier, while a corresponding random number less than 0.2
will not classify it as one.

\section{The methodology of the Probabilistic IAAFT surrogates} \label{sec:sec3}

The SDA methodology focuses on producing simulated sets from a
sequence which capture only the linear properties of the original
data. Then a discriminating pivotal statistic is chosen. Sufficient
evidence for rejecting the null of linear stochastic dynamics is
given when the value of the statistic calculated on the original
data, differs significantly from its values obtained from the
surrogate sets. The simulation procedures for generating surrogate
data differ according to the null being considered. For example, a
simple reshuffling of the original sequence can test for white
noise, whereas more complicated reshuffling exercises may test for
linearly filtered noise or monotonic nonlinear transformations of
linearly filtered noise. Usually the last case is regarded as the
most interesting, as the other procedures may produce spurious
results in the presence of linearly correlated noise that has been
transformed by a static, monotone nonlinearity. The SDA technique is
different to the Bootstrap \cite{bootstrap} as is refers to a
constrained randomization simulation based hypothesis testing
framework, found in permutation tests \cite{Theiler:96}.

To test for the original sequence being a monotonic nonlinear
transformation of linearly filtered noise, one has to simulate
surrogates according to the following steps
\cite{TheilerETAL:92,SCHREIBER:00}:

\begin{enumerate}
 \item Starting with the original sequence $x$, generate an individually and identically distributed
 (i.i.d.)  Gaussian data set $y$ and reorder according to the ranking of $x_n$.
 In this way we can rescale the original sequence to a normal distribution.
 \item Produce the Fourier transform of the rescaled sequence $y$
 and assign a random phase to each (positive) frequency.
 \item Take the inverse transform of above step's sequence,
 say $y^*$. This stage ensures that the surrogates will exhibit the same power
 spectrum as the originating sequence $x$.
   \item Reorder the original data $x$ to generate a surrogate
 $x_s$ which will have  the same rank distribution as $y^*$. In this way we
 are certain that not only the spectrum but also the
 distribution of the original sequence $x$ is preserved in $x_s$.
\end{enumerate} The above  surrogates are referred to as ``\emph{Amplitude
Adjusted Fourier Transformed}" surrogates or AAFT for sort. AAFT
surrogates will have the same distributions and amplitudes with the
original sequence but will not exhibit the same power spectra. To
achieve the latter, an improved, iterative version of AAFT
surrogates (termed IAAFT) has been proposed. To produce IAAFT
surrogates \cite{IAAFT,Kugium:99} one has to follow the steps below:

\begin{enumerate}
   \item Apply a Fourier transform to the original sequence $x$ and
   save the amplitudes $\alpha$. Produce a shuffled surrogate sequence $x_s'$
   from the original  $x$,  apply a Fourier transform to $x_s'$
    and preserve the phases $\phi$. Finally, construct a vector $\vec{r}$
    that contains the ranking of $x$.
   \item Produce a phase randomized (AAFT) surrogate sequence $x_s''$ combining $\alpha$ and $\phi$.
   Compare the rank orders of $x_s''$ and $\vec{r}$. If these are the same,
   proceed to the next step, otherwise the vector $\vec{r}$ hosts the rankings of $x_s''$,
   $\phi$ hosts the phases of $x_s''$, and the procedure of this step is repeated.
   This step can also be terminated if the maximum number of iterations
   defined by the user (e.g., 1000) is reached. Thus we
   avoid strong discrepancies between the surrogates and the
   original sequence's spectrum.
   \item Force $x_s''$ to follow the distribution of $x$,
   by assigning on its indices the corresponding values of $x$.
\end{enumerate}

The IAAFT surrogates ensure that the main linear features of a time
series will be faithfully preserved. However, the above procedure
has been designed for stationary time series and therefore cannot
cope with the presence of heteroscedasticity and outliers. In other
words and with respect to the classification produced in section
\ref{eq:eq2}, the IAAFT surrogates have been designed for time
series where all the observations are subject to $|\mathrm{z}_{
\mathrm{LMS} }| \leq 2 $. According to the proposed framework in
this paper and in order to take into account the outliers that are
observed in stock returns, we have to modify the surrogate
generating algorithm according to the following steps:

\begin{enumerate}
\item Calculate the LMS location parameter of the time series.
\item Calculate the LMS scale parameter of the time series.
\item Calculate the $\mathrm{z}_{ \mathrm{LMS} }$ for each observation.
\item Convert the $\mathrm{z}_{ \mathrm{LMS} }$  to $\lambda$, according to section \ref{sec:sec2}.
\item Create a new series of uniformly distributed random numbers in
[0,1], say  $u$, with length equal to the length of the original
time series.
\item Create a new time series $x_s$, which contains all
the values of $x$ that correspond to $\lambda_i \geq u_i$.
\item Apply the IAAFT surrogate algorithm to $x_s$.
\item The final surrogate sequence will preserve the values of $x$ that correspond to $\lambda_i <
u_i$, in exactly the same positions as in the original sequence, and
will receive the surrogate of $x_s$ for $\lambda_i \geq u_i$, to
fill the remaining gaps.
\end{enumerate}

Our experiments below show that according to the above procedure
(termed Probabilistic IAAFT, or PIAAFT for short), the outliers,
volatility clustering and hence heteroscedasticity can be faithfully
reproduced with a ``reasonable" probability, according to their
level of presence in the original sequence. Moreover, the rest of
the desirable properties of the IAAFT surrogates are preserved.

\section{Calendar Correction} \label{sec:sec4}

So far we have described a surrogates generation procedure which is
able to account for heteroscedasticity. In this section we also
demonstrate  how to account for the calendar effects. As a first
step we have to define what we imply here by the term ``calendar
effects". Since there is no universal definition, we presume eight
kinds of calendar effects. The first five effects, and the least
important ones, are the five weekdays. Next and of greater
importance, the first and last trading days of a month
(day-of-month) are being considered as calendar effects. Finally, we
have the holiday effect, which is also assumed here to be the most
important. For example, if a trading day can be characterized as
both a pre-holiday and end-of-month day, the holiday effect applies.
Following the same rationale, if a trading day is both a Thursday
and the first day of a trading month, it is classified according to
the latter effect.

In order to specialize the algorithm given in section
\ref{sec:sec3}, we have to reconsider its first 3 steps for the
``calendar-wise" time series. To achieve it, we normalize (using the
LMS parameters) every calendar-wise distribution. The rest of the
steps are followed without any change, save for the 7th step which
has to be adapted according to the calendar structure of the time
series. This procedure is the Calendar Corrected version of the
PIAAFT (hence CCPIAAFT).

\section{Empirical Results}

This section compares the surrogates produced by the proposed
CCPIAFFT algorithm to the surrogates of the IAAFT algorithm. Our
time series is the NASDAQ Composite Index, daily closings, from
5-Feb-1971 to 31-Dec-2003. There are totally 8311 observations.
Since all the surrogates generating algorithms need the original
time series to be at least mean stationary, we  work with the first
logarithmic differences of the daily closing prices (i.e., the
continuously compounded returns).

\begin{landscape}
\begin{figure}[ht]
  \centering
  \includegraphics[height=5in]{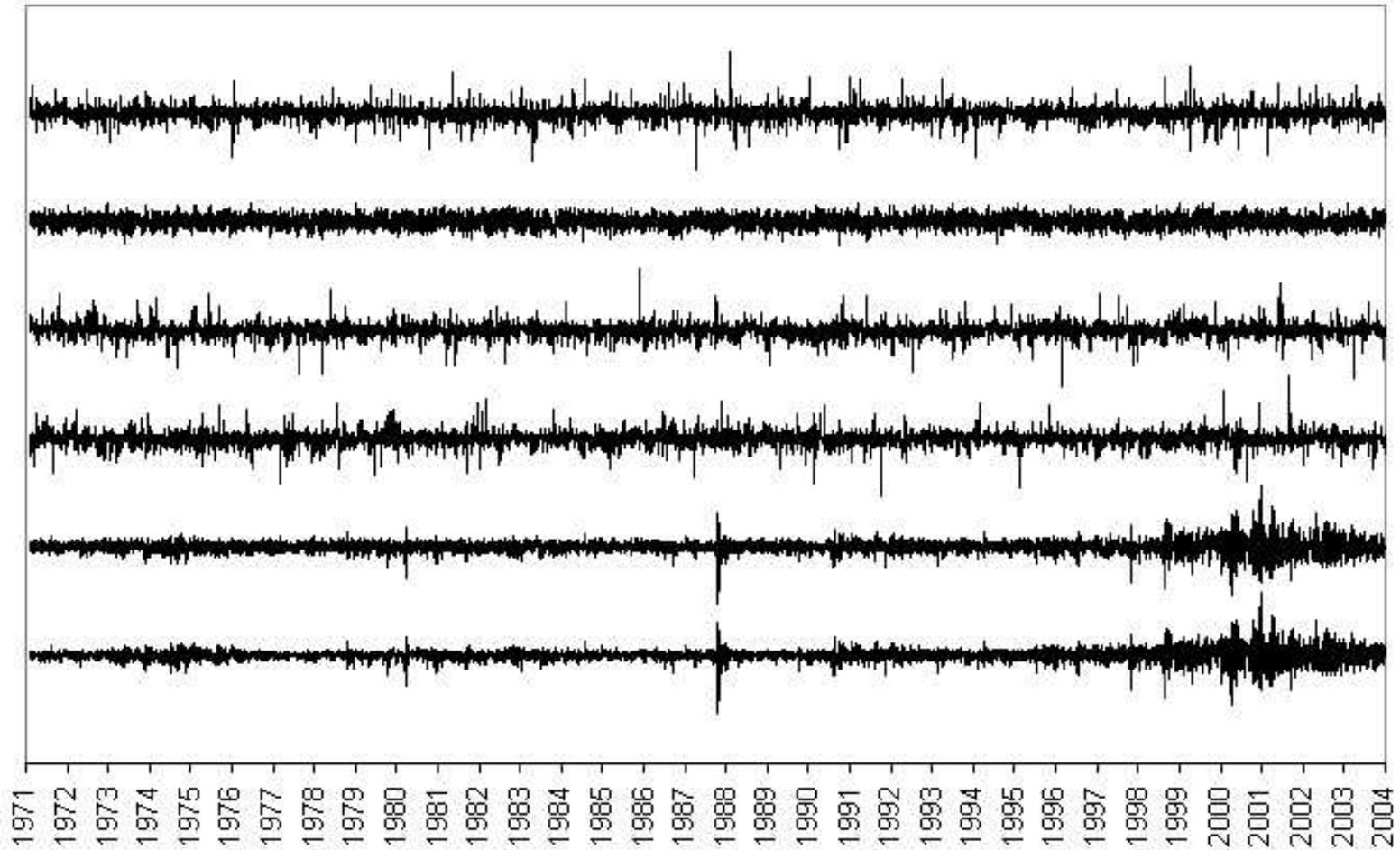}\\
  \caption{The original time series (bottom) and 5 surrogates (from top to bottom):
  the shuffled surrogates (top), the phase randomized surrogates, the AAFT surrogates,
  the  IAAFT surrogates and the CCPIAAFT surrogates. It is evident that the CCPIAAFT series
  preserve the salient features of the original sequence, especially the volatility
  clustering and the outliers (shocks) which are linked to well known historical events such as the crash
  of 1987 and the uncertainty after the burst of the more recent financial bubble.}\label{fig:fig2}
\end{figure}
\end{landscape}

\begin{landscape}
\begin{figure}[ht]
  \centering
  \includegraphics[height=5in]{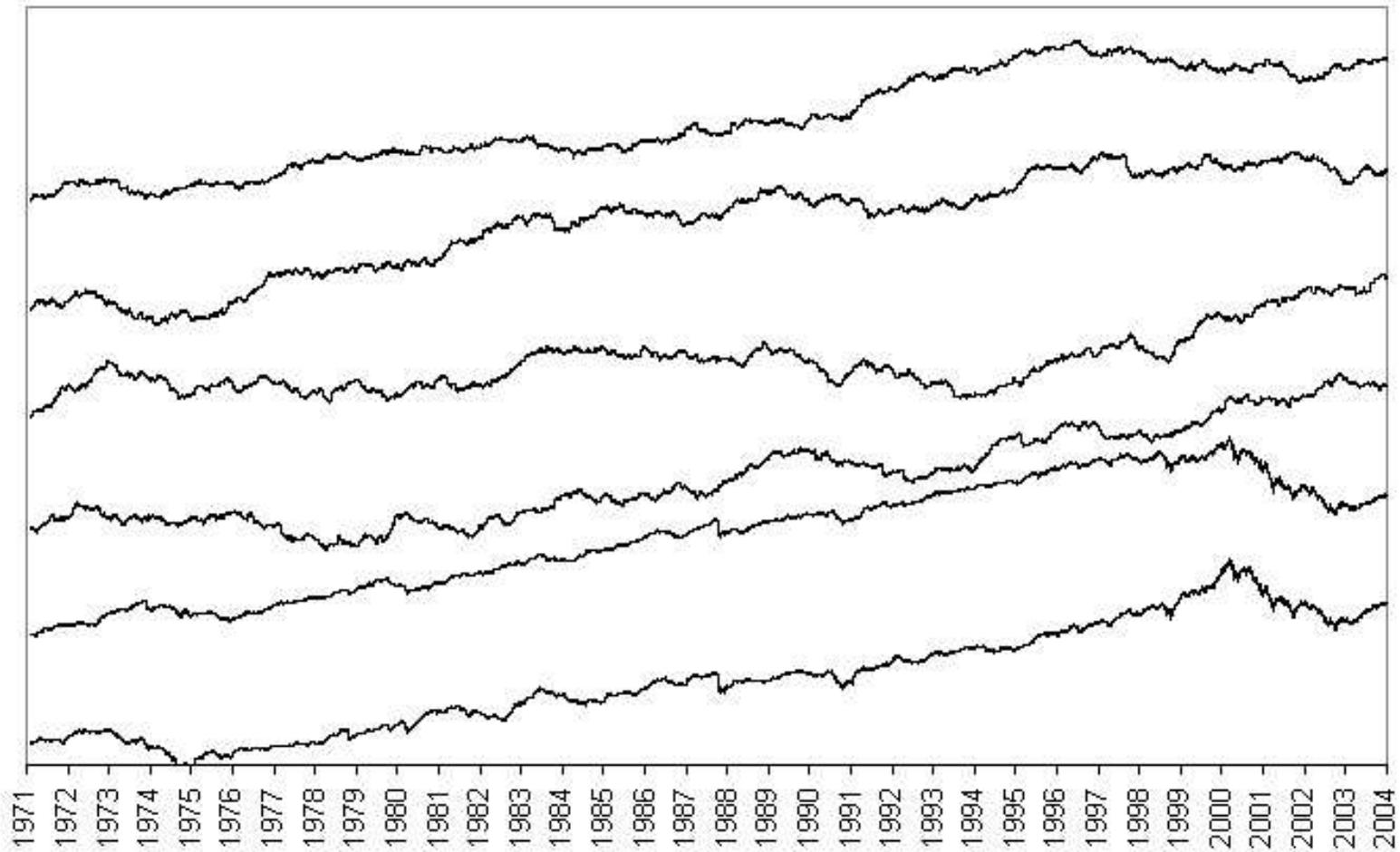}\\
  \caption{The levels of the time series shown in Fig. (\ref{fig:fig2}).
  The CCPIAAFT surrogate series levels (2nd from bottom) preserve exactly the trends
  that the original time series exhibit, while the all the other
  sequences above follow a general trend with no time-specific characteristics.}\label{fig:fig3}
\end{figure}
\end{landscape}

\bigskip
{\begin{center}  [ Insert Fig. \ref{fig:fig2} about here. ]
\end{center}}
{\begin{center}  [ Insert Fig. \ref{fig:fig3} about here. ]
\end{center}}
\bigskip

\begin{figure}[ht]
  \centering
    \subfigure[$m=2$]{\includegraphics*[height=3in]{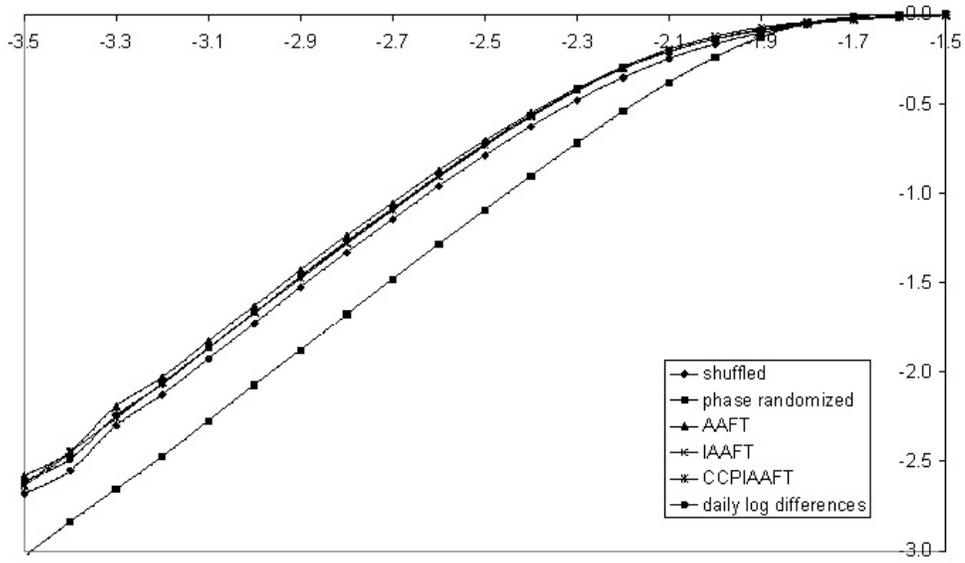}}\\
    \subfigure[$m=3$]{\includegraphics*[height=3in]{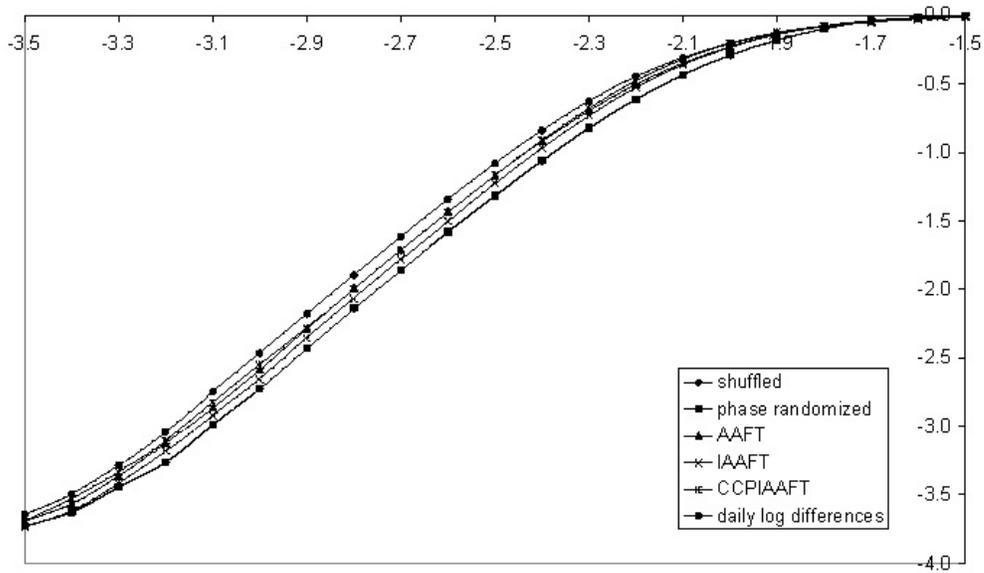}}\\
  \caption{The correlation integral on the series of Fig. (\ref{fig:fig2}) with embedded dimensions
  (a)  $m=2$ and (b) $m=3$. }\label{fig:fig4}
\end{figure}

\begin{landscape}
\begin{figure}[ht]
  \centering
  \includegraphics[height=5in]{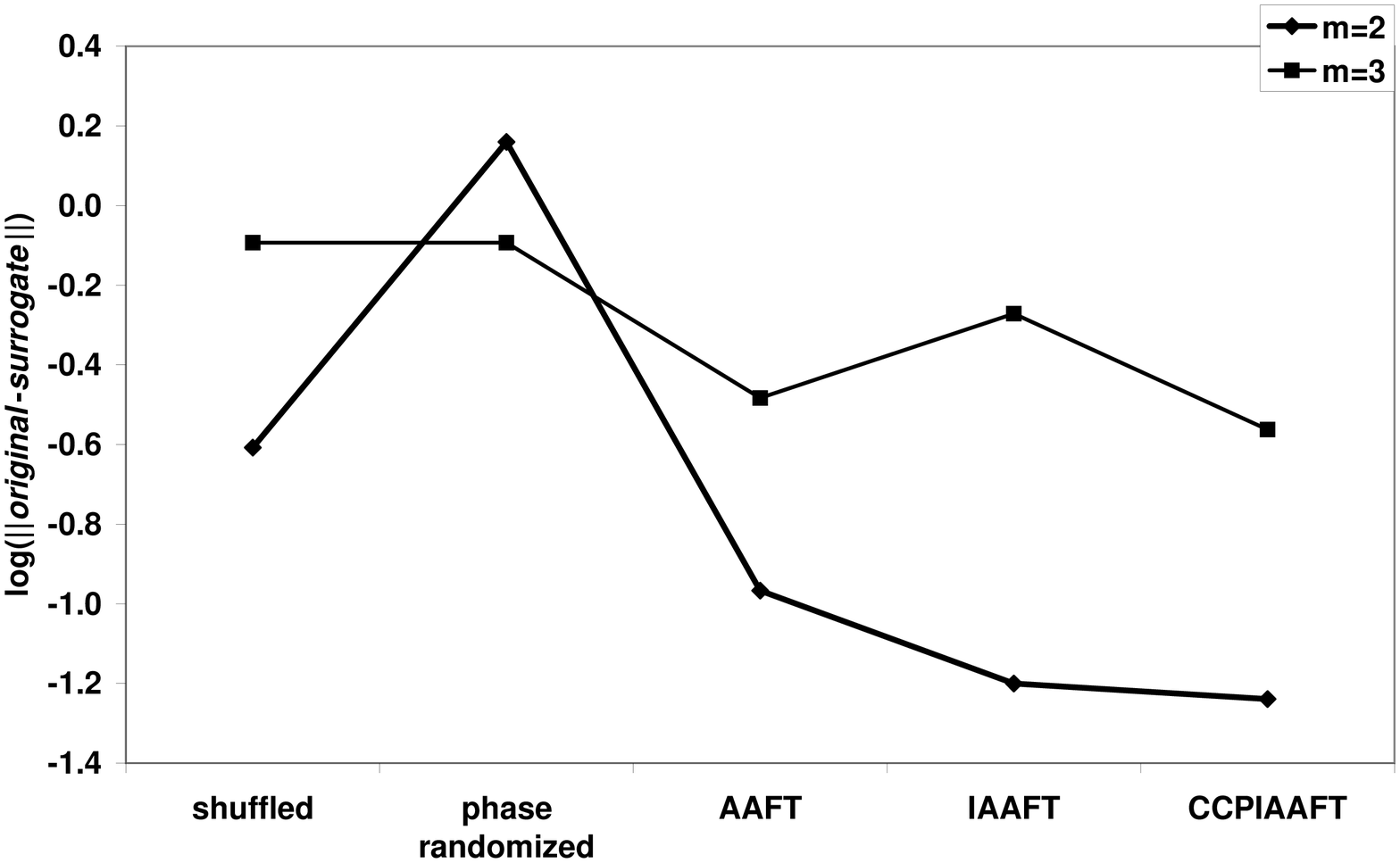}\\
  \caption{The logarithm of the norm-2 difference between the
  correlation integral of the original time series and the surrogates,
  shown in Fig. (\ref{fig:fig2}). We observe that in both cases the CCPIAAFT
  surrogates show the smallest difference compared to their
  counterparts, implying that the CCPIAAFT surrogates provide improved simulations
  of the original time series.}\label{fig:fig5}
\end{figure}
\end{landscape}

As the Fig. \ref{fig:fig2} and \ref{fig:fig3} show, there is no need
for specific statistical tests to realize the difference between the
compared surrogate algorithms. The CCPIAAFT surrogates ``imitate"
extremely well the heteroscedasticity caused by volatility
clustering in the original time series and the trend changes that
are implied. In Fig. \ref{fig:fig4} and \ref{fig:fig5} we utilize
the correlation integral \cite[CI:]{Grassberger83d}  to demonstrate
that the CCPIAAFT surrogates result a CI much more closer to the one
of the original time series.

\bigskip
{\begin{center}  [ Insert Fig. \ref{fig:fig4} about here. ]
\end{center}}
{\begin{center}  [ Insert Fig. \ref{fig:fig5} about here. ]
\end{center}}
\bigskip

\begin{landscape}
\begin{figure}[ht]
  \centering
  \includegraphics[height=5in]{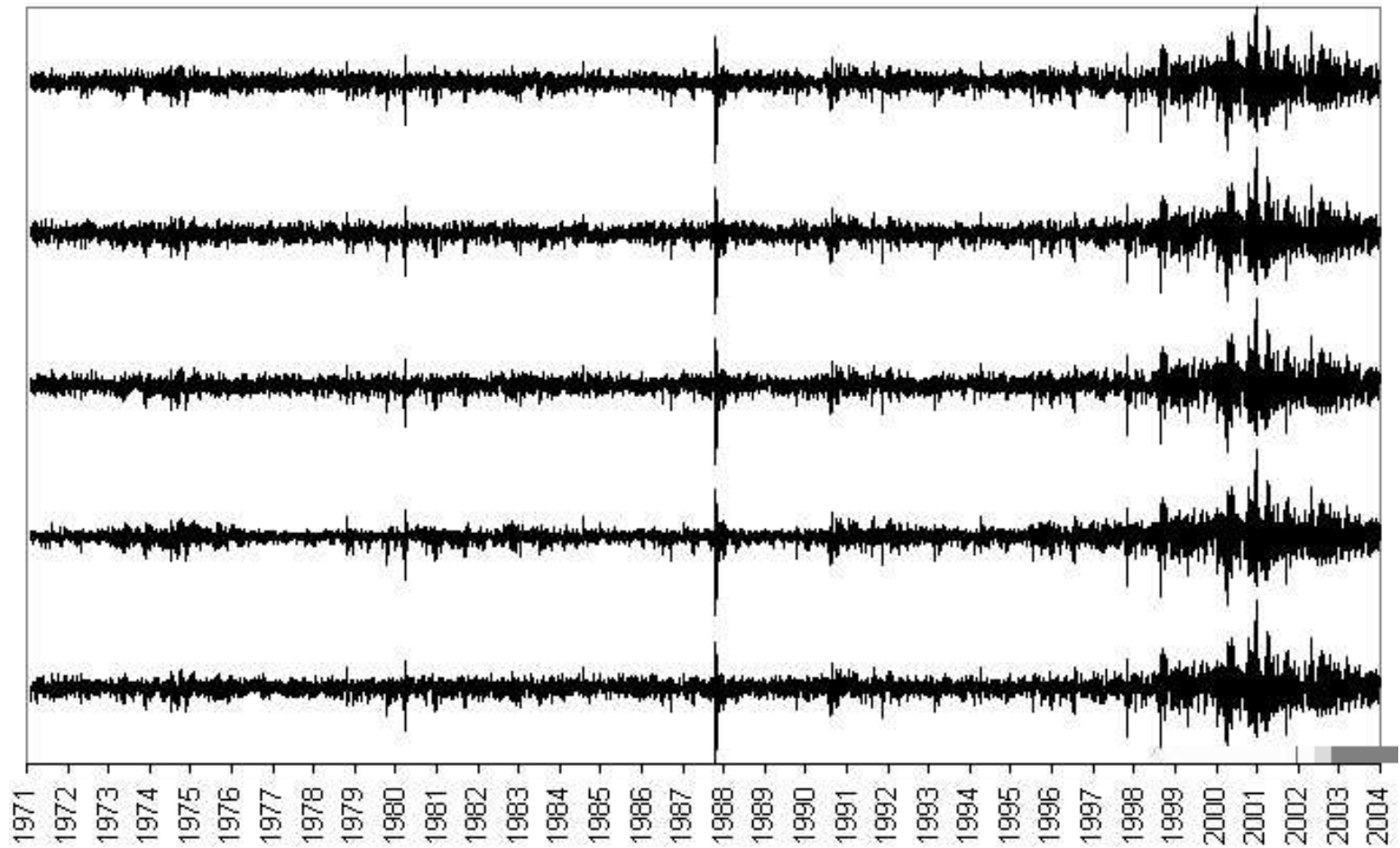}\\
  \caption{A comparison of  the original time series and
   4 CCPIAAFT surrogate series. Which one is the original? (Answer: the 4th from above).
   }\label{fig:fig6}
\end{figure}
\end{landscape}

\begin{landscape}
\begin{figure}[ht]
  \centering
  \includegraphics[height=5in]{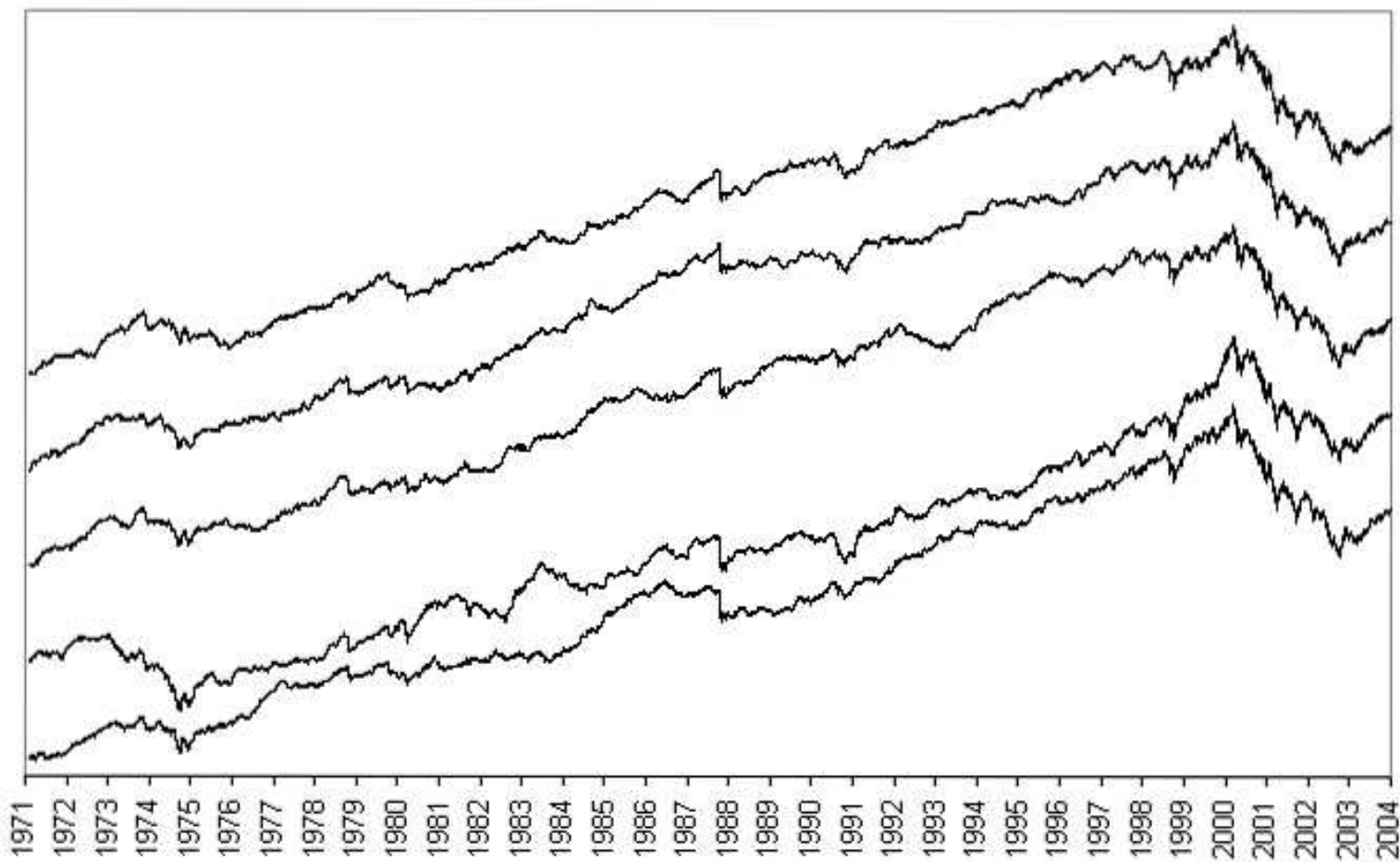}\\
  \caption{The levels of the series shown in Fig. (\ref{fig:fig6}). In this graph the
  differentiation from the original time series is obvious in very few specific
  time domains. More precisely, we can observe that the drop of the index related to the
  1974 crisis and the increase related to the 2000 bubble, appear to be smoother in all
  surrogate series. This is attributed to the small daily changes in each case
  being considered as part of the normal fluctuations of the original time series
  by the CCPIAAFT procedure.}\label{fig:fig7}
\end{figure}
\end{landscape}

Considering the IAAFT surrogates as our null hypothesis implies that
we theorize that extreme events (such as the oil crisis of 1973, the
Black Monday of 1987 and the recent bubble of 2000) can occur with
equal probability, a premise that voluminous research in finance has
challenged so far. Certain events that trigger unanticipated
changes, occur due to exogenous political and economic (and not
necessarily) market dynamics. Therefore, if these unsystematic
fluctuations could be preserved, along with any other calendar
effects, one could produce financial surrogates that faithfully
reproduce certain market realities. The linear correlations and the
randomization of the returns should only affect the systematic
components. Hence, CCPIAAFT surrogates  essentially isolate the
systematic from the unsystematic changes. The degree to which this
is achieved is highlighted in Fig. \ref{fig:fig2} and
\ref{fig:fig3}. Fig. \ref{fig:fig6} and \ref{fig:fig7} also refer to
various realizations of CCPIAAFT surrogates for comparison purposes.

\bigskip
{\begin{center}  [ Insert Fig. \ref{fig:fig6} about here. ]
\end{center}}
{\begin{center}  [ Insert Fig. \ref{fig:fig7} about here. ]
\end{center}}
\bigskip

\section{Conclusions\label{sec:conclusions}}

In this paper we suggest a method which embodies the outliers and
calendar effects on the production of surrogate data. In financial
time series where heteroscedasticity, in the sense of volatility
clustering, is the most striking feature, the proposed method yields
simulated sequences which are more similar to the original time
series, when compared with other surrogate data generating methods.
Moreover, the proposed approach has the advantage of behaving as the
IAAFT algorithm when no heteroscedasticity or calendar effects are
present. We do not assume (G)ARCH volatility structures, however our
strategy can be modified to accommodate such a case. We reserve this
as an area for future research.


\begin{tiny}

\end{tiny}

\clearpage \pagestyle{empty}

\clearpage

\clearpage

\clearpage

\clearpage

\clearpage

\clearpage

\end{document}